\documentclass[aps,prd,reprint,groupedaddress]{revtex4-1}

\usepackage{graphics} 
\usepackage{amsbsy}

\begin{document}


\title{Radiative effects in the processes of exclusive photon electroproduction from polarized protons }


\author{Igor Akushevich}
\email[]{igor.akushevich@duke.edu}
\affiliation{Physics Department, Duke University, Durham, NC 27708, USA\\ and \\ Jefferson Lab., Newport News, VA 23606, USA}
\author{Alexander Ilyichev}
\affiliation{National Center for Particle and High Energy Physics, Byelorussian State University, Minsk, 220040, Belarus}

\date{\today}

\begin{abstract}
Radiative effects in the electroproduction of photons in  polarized $ep$-scattering are calculated in the leading log approximation and analyzed numerically for kinematical conditions of current measurement at Jefferson Lab. Radiative corrections to the cross sections, their azimuthal distributions and Fourier coefficients are in particular focus. Kinematical regions where the radiative corrections are considerable are identified.  
\end{abstract}


\maketitle
\section{\label{Intro}Introduction}

The processes of the photon electroproduction  are intensively investigated both theoretically \cite{BelitskyMuller2009PRD,BeMu2010PRD} and experimentally \cite{Camacho_etal_2006_PRL,Mazouz_etal_2007_PRL,Girod_etal_2008_PRL}. The cross section of the process is sensitive to the deep virtual Compton scattering (DVCS) amplitude that is of great interest due to its connection to generalized parton distributions.  
The Bethe-Heitler (BH) process
is not distinguishable from the DVCS measurements and therefore it is   
the basic background contribution to the observed cross section.  
One obstacle in the analysis of vast data on DVCS collected in Jlab experiments is the deficit of comprehensive theoretical calculations of radiative corrections (RC) including the effects of hard photon emission with controlled accuracy. Available calculations of QED radiative effects in \cite{AkBaSh1986YP,Vanderhaeghen2000,ByKuTo2008PRC,AkKuSh2000PRDtail,AkKuSh2001PRDDVCS} have certain limitations and cannot cover all modern requirements of experimental data analysis on  photon electroproduction.   In this paper we present the radiative correction calculations to the cross section of BH in leading approximation. The main process contributing to the RC is two-photon emission, i.e., $e+p\rightarrow e'+p'+2\gamma$. Another contribution is due to one-loop effects in $e+p\rightarrow e'+p'+\gamma$.
In the approximation the only leading term containing $L=\log(Q^2/m^2)$ ($m$ is the electron mass) is kept. For Jlab kinematics $L\sim 15$ and therefore, the used approximation allows to keep the major part of the RC.

The paper is organized as follows. The BH cross section is calculated in Section \ref{Born}. Specific attention is paid on explicit representation of the BH cross section including polarization part of the cross section and mass corrections, as well as for angular structure of the BH cross section. RC calculation is performed in Section \ref{RC}. First we calculate the matrix 
element squared and trace all sources of occurrence of the electron mass dependence. Second, we represent the phase space of two final photons, introduce the so-called shifted kinematics, and calculate integrals over additional photon phase space. Third, we add the contribution of loops and calculate the lowest order RC to the BH cross section. Fourth we generalize the result for the RC to the BH cross section to represent the higher order corrections. Section \ref{SectNumeric} presents the numeric estimates of the radiative effects in current experiments at JLab focusing on the RC to cross section in a wide kinematic region and angular structure (i.e., respective Fourier coefficients) with specific focus on the coefficient not appeared in the BH cross section but generated by RC.             
Finally in Section \ref{SectDiscussion} 
we discuss the most interesting features of our findings in the theoretical calculation and numeric results (e.g., importance of mass corrections and kinematical regions with large effects generated by RC), briefly describe the state-of-art in calculations of RC to exclusive photon electroproduction processes and the place of our calculation among other calculations, and comment perspectives in further theoretical development in RC in exclusive photon electroproduction measurements.  

\section{\label{Born}The BH cross sections}

The BH process 
\begin{equation}\label{BHprocess}
e(k_1)+p(p)\longrightarrow e'(k_2)+p'(p')+\gamma(k),
\end{equation}
is traditionally described using four kinematical variables: $Q^2=-(k_1-k_2)^2$, $x=Q^2/(2p(k_1-k_2))$, $t=(p-p')^2$, and $\phi$, the angle between planes 
$({\bf k_1},{\bf k_2})$ and $({\bf q},{\bf p'})$ ($q=k_1-k_2$).

\begin{figure}[b]\centering
\scalebox{0.32}{\includegraphics{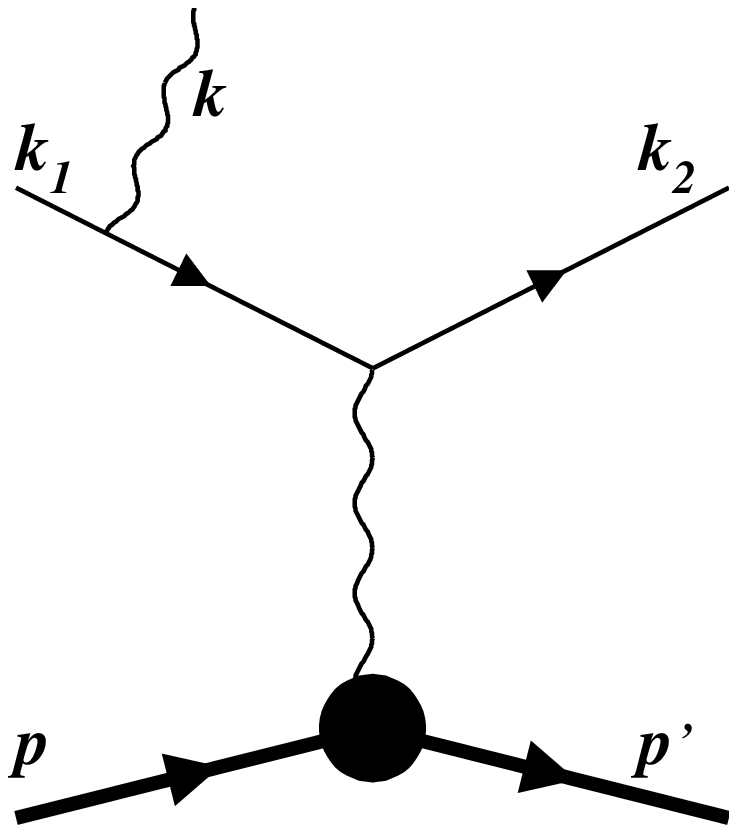}}
\hspace{0.4cm}
\scalebox{0.32}{\includegraphics{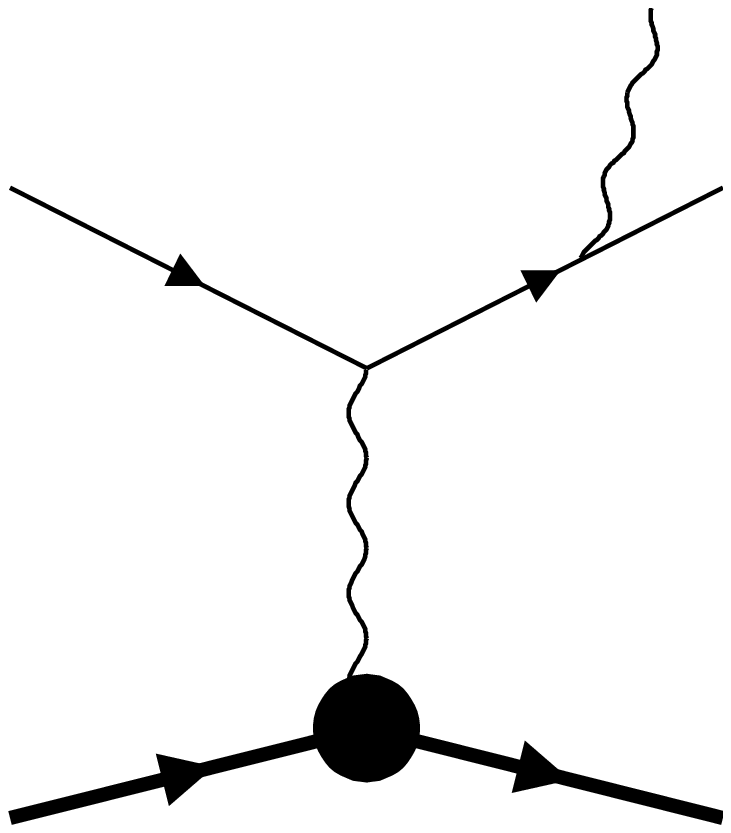}}
\\
{\bf a) \hspace{2cm} b)}
\caption{\label{BHgraphs}Feynman graphs of BH process}
 \end{figure}

The BH matrix element is 
${\cal M}_{BH}=e^3t^{-1}J^h_\mu J_{\mu}^{BH}$ with 
\begin{equation}
J^h_\mu={\bar u}(p')\biggl(\gamma_{\mu}F_1+i\sigma_{\mu\nu}\frac{p_\nu '-p_\nu}{2M}F_2\biggr)u(p)
\end{equation}
and
\begin{eqnarray}
J_{\mu}^{BH}&=& 
{\bar u}_2\Biggl [
 \gamma_\mu \frac{{\hat k}_1-{\hat k}+m}{-2kk_1}{\hat \epsilon}
+ {\hat \epsilon} \frac{{\hat k}_2+{\hat k}+m}{2kk_2}\gamma_\mu 
\Biggr ]u_1
\nonumber \\&=& -
{\bar u}_2\Biggl [\left(\frac {k_1\epsilon}{kk_1}-\frac {k_2\epsilon}{kk_2}\right)\gamma_\mu
-\frac{\gamma_\mu \hat{k}\hat{\epsilon}}{2kk_1}
-\frac{\hat{\epsilon}\hat{k}\gamma_\mu }{2kk_2}
\Biggr ]u_1, 
\end{eqnarray}
where ${\bar u}_2\equiv {\bar u}(k_2)$, ${u}_1\equiv {u}(k_1)$, and $\epsilon$ is the photon polarization vector. 
The matrix element ${\cal M}_{BH}$ corresponds to the graphs in Figure 
\ref{BHgraphs}a and \ref{BHgraphs}b.

The cross section of the BH process is
\begin{eqnarray}\label{dGamma}
d\sigma_0&=&\frac 1{2S} {\cal M}_{BH}^2 d\Gamma_0
=
\frac{32\pi^3\alpha^3}{St^2} \bigl(J^h_\mu J^{BH}_{\mu}\bigr)^2 d\Gamma_0,
\end{eqnarray}
where $S=2ME_1$, $E_1$ is the beam energy in the lab. system, and $M$ is the proton mass.

 Phase space for the BH cross section is parametrized as 
\begin{eqnarray}
d\Gamma_0&=&\frac 1{(2\pi)^5}
\frac{d^3k_2}{2E_2}
\frac{d^3p'}{2p'_0}
\frac{d^3k}{2\omega}
\delta^4(k_1+p-k_2-p'-k)
\nonumber \\&=&
\frac{Q^2dQ^2dxdtd\phi}{(4\pi)^4x^2 S \sqrt{\lambda _Y}}
\end{eqnarray}
with $\lambda _Y=S_x^2+4 M^2Q^2$ and $S_x=S-X=Q^2/x$. Kinematical limits on $t$ are defined as
\begin{eqnarray}
t_{2,1}=\frac{-1}{2W^2}\bigl((S_x-Q^2)(S_x\pm\sqrt{\lambda_Y})+2M^2Q^2\bigr),
\end{eqnarray}
where $W^2=S_x-Q^2+M^2$.

The 4-fold BH cross section ($\sigma_{BH}\equiv d\sigma_0/dQ^2dxdtd\phi$) including both unpolarized and spin dependent parts is
\begin{equation}\label{BHCS}
\sigma_{BH}= { -\alpha^3 Q^2 \over 4\pi S^2 x^2 t \sqrt{\lambda_Y}}\sum_{i=1}^4 (T_i+2m^2\hat{T}_i){\cal F}_i.
\end{equation}
Those terms proportional to the lepton mass squared ($m^2$) are kept that give finite (i.e., non-vanishing  for $m\rightarrow 0$) contribution after integration over $\phi$. First two terms, ($(T_1+2m^2\hat{T}_1){\cal F}_1$ and $(T_2+2m^2\hat{T}_2){\cal F}_2$), describe unpolarized cross section and last two terms, ($(T_3+2m^2\hat{T}_3)){\cal F}_3$ and $(T_4+2m^2\hat{T}_4){\cal F}_4$), correspond to the spin dependent part of the cross section. The quantities ${\cal F}_i$ are squared combinations of nucleon formfactors: 
\begin{eqnarray}
{\cal F}_1&=&{\cal F}_3=(F_1(t)+F_2(t))^2,
\nonumber \\
{\cal F}_2&=&\frac{4}{t}(F_1^2(t)+\frac{-t}{4M^2}F_2^2(t)),
\nonumber \\
{\cal F}_4&=&\frac{4}{t}(F_1(t)+F_2(t))(F_1(t)+\frac{t}{4M^2}F_2(t)).
\end{eqnarray}
The quantities $T_i$ are 
\begin{eqnarray}
T_1&=&\frac{2}{u_0w_0}(u_0^2 + w_0^2 - 2Q^2t ),
\nonumber \\
T_2&=&\frac{M^2}{2}T_1+\frac{t}{u_0w_0}(S^2+X^2 \nonumber\\&&\qquad
-Q^2S_x-Sw_0-Xu_0),
\nonumber \\
T_3&=&\frac{4Mt(\eta p')}{\lambda_tu_0w_0}
(2X(u_0-Q^2)-2S(w_0+Q^2)
\nonumber\\&&\qquad +(u_0 + w_0)(Q^2-t)),
\nonumber \\
T_4&=&-M^2T_3+\frac{2M}{u_0w_0}\Bigl(
(Q^2-u_0)\bigl(t(\eta k_2)+X(\eta p')\bigr) \nonumber\\&&
+(Q^2+w_0)\bigl(t(\eta k_1)+S(\eta p')\bigr)\Bigr),
\end{eqnarray}
where $\lambda_t=t(t-4M^2)$ and $\eta$ is the target polarization vector. 
The quantities represented the lepton mass corrections are
\begin{eqnarray}
\hat{T}_1&=&\frac{2t}{u_0^2} + \frac{2t}{w_0^2},
\nonumber \\
\hat{T}_2&=&\frac{M^2}{2}\hat{T}_1+\frac{S^2+St}{u_0^2}+\frac{X^2+Xt}{ w_0^2},
\nonumber \\
\hat{T}_3&=&\frac{4M(\eta p')}{t-4M^2}\biggl(
(2S+t)\Bigl(\frac{1}{u_0^2} 
+ \frac{S_x}{Sw_0^2}\Bigr)
\nonumber\\&&
+
\frac{1}{w_0^2}\Bigl(-Q^2+\frac{1}{S}(X^2+(X-t)^2) \biggr),
\nonumber \\
\hat{T}_4&=&-M^2\hat{T}_3+2M\Bigl(
\frac{S_{x}+t}{Sw_0^2}\bigl(t(\eta k_2)+X(\eta p')\bigr) \nonumber\\&&
-\Bigl( \frac{1}{u_0^2} + \frac{1}{w_0^2}\Bigr)\bigl(t(\eta k_1)+S(\eta p')\bigr)\Bigr).
\end{eqnarray}

All variables including the scalar products $\eta k_1$, $\eta k_2$, and $\eta p'$ are ultimately expressed in terms of 5 kinematical variables: $S$, $t$, $Q^2$, $x$ and $\phi$, e.g.,
\begin{eqnarray}
w_0&=& 2kk_1=-\frac{1}{2}(Q^2+t)
+\frac{S_p}{2\lambda_Y}\bigl(S_x(Q^2-t)+2tQ^2\bigr)+
\nonumber\\&& \qquad \qquad
+\frac{\sqrt{\lambda_{uw}}}{\lambda_{Y}}\cos\phi ,
\nonumber \\
u_0&=& 2kk_2=w_0+Q^2+t
\end{eqnarray}
with $S_p=S+X$ and 
\begin{equation}\label{lambdauw}
\lambda_{uw}=4Q^2W^2(SX-M^2Q^2-m^2\lambda_Y)(t-t_1)(t_2-t).
\end{equation}
An additional azimuthal angle is required to describe the case of transversely polarized target (see eq. (\ref{etatrans})). Note that in massless approximation (for $m\rightarrow 0$) the BH cross section exactly coincides with results of \cite{BKM2002}.  
The following equations relating our notation to the notation of ref. \cite{BKM2002} 
are valid: $u={\cal P}_2Q^2$, $w=-{\cal P}_1Q^2$, and (for $m\rightarrow 0$) $\lambda_{uw}=4Q^4S^2S_x^2K^2$.

Explicit expressions for the scalar products of momenta with target polarization are
\begin{eqnarray}
\eta k_1&=&-{SS_x+2M^2Q^2 \over 2M\sqrt{\lambda_Y}}, 
\nonumber \\
\eta k_2&=&-{XS_x-2M^2Q^2 \over 2M\sqrt{\lambda_Y}}, 
\nonumber \\
\eta p'&=&-{-tS_x+2M^2(Q^2-t) \over 2M\sqrt{\lambda_Y}}
\end{eqnarray}
for longitudinal part of proton polarization vector (i.e., $\boldsymbol \eta$  $|| \bf q$) and 
\begin{eqnarray}\label{etatrans}
\eta k_1&=&-\sqrt{\lambda_{SX} \over \lambda_Y}\cos (\varphi +\phi ), 
\nonumber \\
\eta k_2&=&-\sqrt{\lambda_{SX} \over \lambda_Y}\cos (\varphi +\phi ),
\nonumber \\
\eta p'&=&-{Q^2SS_xK \over \sqrt{\lambda_{SX}\lambda_Y}} \cos \varphi
\end{eqnarray}
for its transverse part (i.e., ${\boldsymbol \eta } \perp {\bf q}$). Here $\lambda_{SX}=SXQ^2-M^2Q^4$ and $\varphi$ is the angle between polarization and production planes, i.e., planes defined as $({\bf q},{ {\boldsymbol \eta }_T})$ and $({\bf q},{\bf p'})$). In the general case ($\eta=(0,\eta_x,\eta_y,\eta_z)$) the scalar products are
\begin{eqnarray}
\label{etaxyz}
\eta k_1&=&-\sqrt{\lambda_{SX} \over \lambda_Y}\;\eta_x-{SS_x+2M^2Q^2 \over 2M\sqrt{\lambda_Y}}\;\eta_z, 
\nonumber \\
\eta k_2&=&-\sqrt{\lambda_{SX} \over \lambda_Y}\;\eta_x-{XS_x-2M^2Q^2 \over 2M\sqrt{\lambda_Y}}\;\eta_z, 
\nonumber \\
\eta p'&=&-{Q^2SS_xK \over \sqrt{\lambda_{SX}\lambda_Y}} (\eta_x\cos\phi+\eta_y\sin\phi)
\nonumber \\&&\quad
-{-tS_x+2M^2(Q^2-t) \over 2M\sqrt{\lambda_Y}}\;\eta_z.
\end{eqnarray}

The cross section $\sigma_{BH}$ is defined as the 4-dimensional cross section in eq. (\ref{dGamma}). It means that integration over $\varphi$ is assumed to be performed resulting in additional factor $2\pi$. This is because of respective symmetry of the process with unpolarized or longitudinally polarized target. The cross section of transversely polarized target explicitly depends on $\varphi$ because of (\ref{etatrans}). Thus $\sigma$ in the case of transversely polarized target means 5-dimensional cross section with addition of $d\varphi/2\pi$.

\begin{figure}\centering
\scalebox{0.24}{\includegraphics{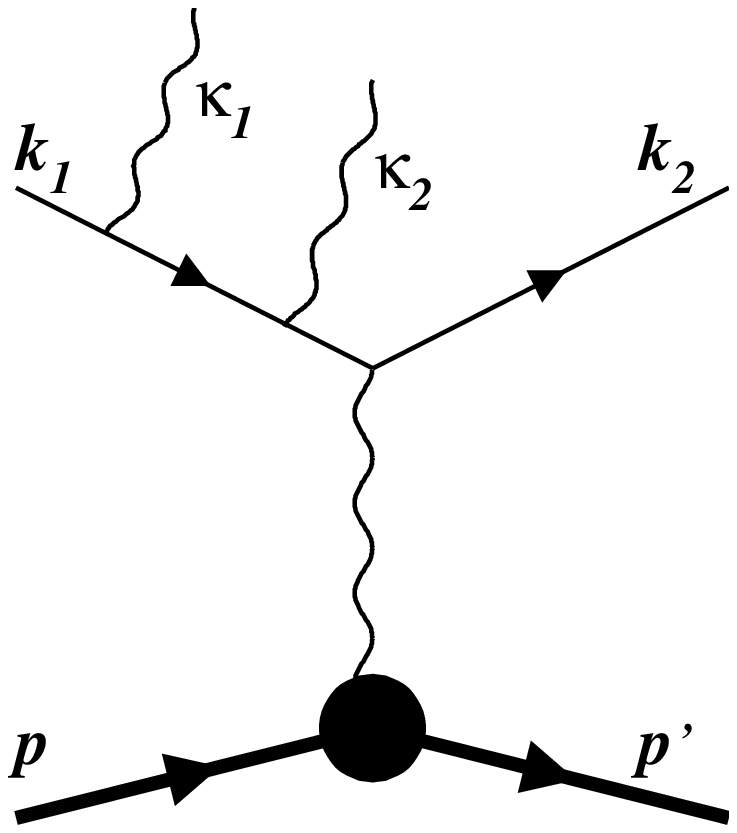}}
\hspace{0.25cm}
\scalebox{0.24}{\includegraphics{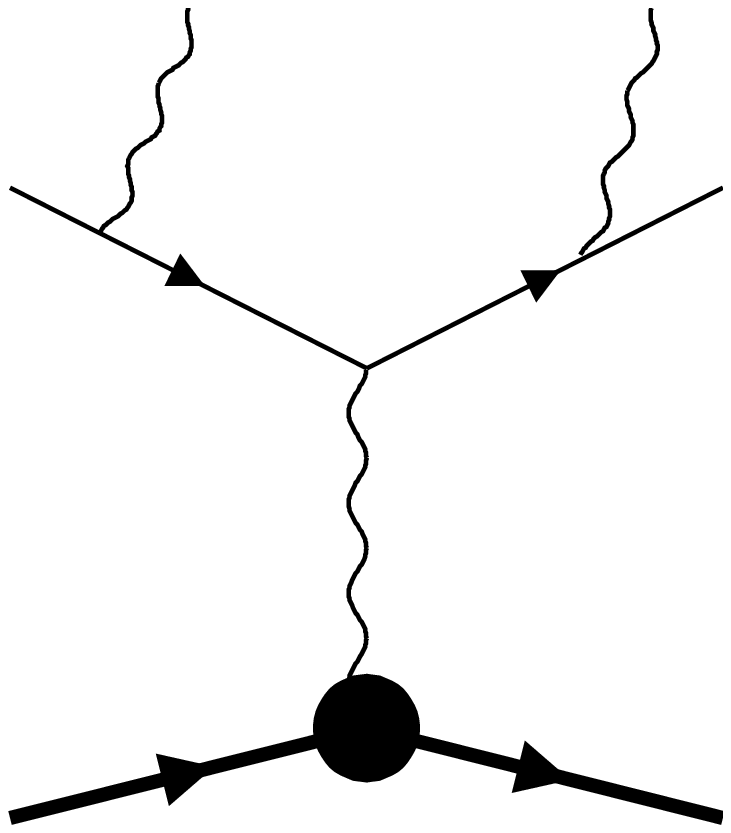}}
\hspace{0.25cm}
\scalebox{0.24}{\includegraphics{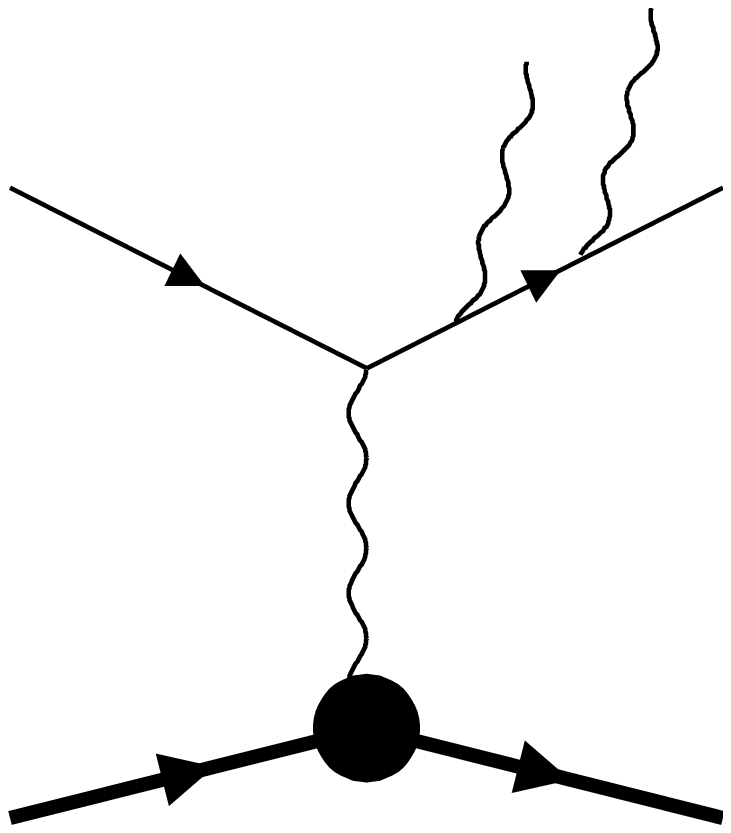}}
\\
{\bf a) \hspace{1.8cm} b) \hspace{1.8cm} c)}
\caption{\label{Twoggraphs}Feynman graphs of two real photon emission}
 \end{figure}

\subsection{\label{polar}Angular structure of the BH cross section}

Azimuthal structure of the BH cross sections are often of interest both theoretically and experimentally. Dependence of the BH cross section (\ref{dGamma}) on the angle $\phi$ appears in $u_0$ and $w_0$ in numerator and denominator of the BH cross section and in scalar products $\eta k_1$ and $\eta k_2$ for transversely polarized proton (angle $\varphi$ is assumed to be fixed, i.e., $\phi$-independent). The upolarized and spin-dependent parts of BH cross section can be presented in the form:
\begin{eqnarray}\label{scc0}
\sigma_{BH}^{unp}&=&\frac{f}{{\cal P}_1{\cal P}_2}(c_{0,unp}+c_{1,unp}\cos\phi+c_{2,unp}\cos 2\phi),
\nonumber \\
\sigma_{BH}^{LP}&=&\frac{f}{{\cal P}_1{\cal P}_2}(c_{0,LP}+c_{1,LP}\cos\phi),
\nonumber \\\label{ccctp}
\sigma_{BH}^{TP}&=&\frac{f}{{\cal P}_1{\cal P}_2}(c_{0,TP}+c_{1,TP}\cos\phi+s_{1,TP}\sin\phi),
\end{eqnarray}
where $f=\alpha^3 S_x^3/(8\pi x^3t\lambda_Y^{5/2})$. The Fourier coefficients  are expressed as
\begin{eqnarray}\label{cc0}
c_0&=&\frac{1}{2\pi f}\int_0^{2\pi}d\phi {\cal P}_1{\cal P}_2\;\;\sigma_{BH}, 
\nonumber \\
c_1&=&\frac{1}{\pi f}\int_0^{2\pi}d\phi \cos\phi\;\;{\cal P}_1{\cal P}_2\;\;\sigma_{BH},
\nonumber \\
c_2&=&\frac{1}{\pi f}\int_0^{2\pi}d\phi \cos2\phi\;\;{\cal P}_1{\cal P}_2\;\;\sigma_{BH},
\nonumber \\
s_1&=&\frac{1}{\pi f}\int_0^{2\pi}d\phi \sin\phi\;\;{\cal P}_1{\cal P}_2\;\;\sigma_{BH} \label{ss1}.
\end{eqnarray}
Only terms $T_1-T_4$ contribute to eq.(\ref{cc0}) while mass corrections represented by ${\hat T}_1-{\hat T}_4$ can be neglected. This is because the integration over $\phi$ in (\ref{cc0}) is performed with weights ${\cal P}_1{\cal P}_2$ reducing singularity level in this terms resulting in their zeroth contribution to (\ref{cc0}) in massless approximation.
The Fourier coefficients (\ref{cc0}) are defined in exactly the same way as those given in eqs. (35-42) of  \cite{BKM2002}. Our analytic calculation of the Fourier coefficients for unpolarizad, longitudinally and transversely polarized cross sections and their subsequent analytical comparison with the expressions of \cite{BKM2002} show that both sets of formulae are identical. Therefore, we do not show the explicit expressions for the Fourier coefficients here. Note, however, that for  
transversely polarized case the Fourier coefficients still depend on $\varphi$ (defined after eq. \ref{etatrans})
 that must be assumed to be fixed to have eqs. (\ref{ccctp}) and (\ref{cc0}) valid. Alternatively, one can assume that the angle between scattering and polarization planes ($\bar\varphi=\phi+\varphi$). In this case $\varphi$-dependence for transversely polarized target needs to be moved out from the expressions for the Fourier coefficients in Eq. (\ref{ccctp}) and that equation needs to be rewritten as 
\begin{eqnarray}\label{barc}
&&\sigma_{BH}^{TP}=\frac{f}{{\cal P}_1{\cal P}_2}({c}^\prime_{0,TP}\cos\varphi+{c}^\prime_{1,TP}\cos\varphi\cos\phi
\\&&\quad
+{s}^\prime_{1,TP}\sin\varphi\sin\phi)=\frac{f}{2{\cal P}_1{\cal P}_2}(
({c}^\prime_{1,TP}-{s}^\prime_{1,TP})\cos{\bar\varphi}
\nonumber\\&&\quad
+2{c}^\prime_{0,TP}\cos(\phi-\bar\varphi)
+({c}^\prime_{1,TP}+{s}^\prime_{1,TP})\cos(2\phi-\bar\varphi)
)=\nonumber
\\&&\quad
=\frac{f}{{\cal P}_1{\cal P}_2}({\bar c}^{}_{0,TP}\cos{\bar\varphi}+{\bar c}^{}_{1,TP}\cos(\phi-\varphi)
\nonumber
\\&&\qquad \qquad \qquad \qquad \qquad\qquad \qquad
+{\bar c}^{}_{2,TP}\cos(2\phi-\varphi)).\nonumber 
\end{eqnarray}

 Important research questions are what magnitude of RC for Fourier coefficients  is and whether RC can generate new functions (e.g., involving $\cos3\phi$ or $\sin 2\phi$) vanishing at the level of the BH cross section.

\section{\label{RC}RC cross section}

The cross section of two photon emission, i.e., the process
\begin{equation}\label{twogammaprocess}
e(k_1)+p(p)\longrightarrow e'(k_2)+p'(p')+\gamma(\kappa_1)+\gamma(\kappa_2),
\end{equation}
is
\begin{eqnarray}
d\sigma&=&\frac{1}{4S} \biggl(\sum_{i=1}^6{\cal M}_i\biggr)^2 d\Gamma,
\end{eqnarray}
where additional factor 2 in the denominator is because there are two identical particles (photons) in the final state. Phase space is parametrized as:   
\begin{eqnarray}
d\Gamma&=&\frac 1{(2\pi )^8}
\frac{d^3k_2}{2E_2}
\frac{d^3p'}{2p'_0}
\frac{d^3\kappa_1}{2\omega_1}
\frac{d^3\kappa_2}{2\omega_2}
\nonumber \\&&\times
\delta^4(k_1+p-k_2-p'-\kappa_1-\kappa_2).
\end{eqnarray}
 
Six matrix elements of the process with emission of additional photon correspondent 
to graphs in Figure \ref{Twoggraphs} are denoted ${\cal M}_{1-6}=e^4t^{-1}J_\mu^h J_{1-6,\mu}$. The quntities $J_{1-6,\mu}$, proportional to the leptonic currents, are:
\begin{eqnarray}
J_{1\mu}&=& 
{\bar u}_2 
\gamma_\mu 
\frac{{\hat k}_1-{\hat \kappa}+m}{-2\kappa k_1+V^2}
{\hat \epsilon}_2
\frac{{\hat k}_1-{\hat \kappa}_1+m}{-2k_1\kappa_1}
{\hat \epsilon}_1
u_1,
\nonumber \\
J_{2\mu}&=& 
{\bar u}_2 
\gamma_\mu 
\frac{{\hat k}_1-{\hat \kappa}+m}{-2\kappa k_1+V^2}
{\hat \epsilon}_1
\frac{{\hat k}_1-{\hat \kappa}_2+m}
{-2k_1\kappa_2}{\hat \epsilon}_2
u_1,
\nonumber \\
J_{3\mu}&=& 
{\bar u}_2 
{\hat \epsilon}_2
\frac{{\hat k}_2+{\hat \kappa_2}+m}{2k_2\kappa_2}
{\hat \epsilon}_1
\frac{{\hat k}_2+{\hat \kappa }+m}{2\kappa k_2+V^2}
\gamma_\mu 
u_1,
\nonumber \\
J_{4\mu}&=& 
{\bar u}_2 
{\hat \epsilon}_1
\frac{{\hat k}_2+{\hat \kappa_1}+m}{2k_2\kappa_1}
{\hat \epsilon}_2
\frac{{\hat k}_2+{\hat \kappa }+m}{2\kappa k_2+V^2}
\gamma_\mu 
u_1,
\nonumber \\
J_{5\mu}&=& 
{\bar u}_2 
{\hat \epsilon}_1
\frac{{\hat k}_2+{\hat \kappa_1}+m}{2k_2\kappa_1}
\gamma_\mu 
\frac{{\hat k}_1-{\hat \kappa}_2+m}{-2k_1\kappa_2}
{\hat \epsilon}_2
u_1,
\nonumber \\
J_{6\mu}&=& 
{\bar u}_2 
{\hat \epsilon}_2
\frac{{\hat k}_2+{\hat \kappa_2}+m}{2k_2\kappa_2}
\gamma_\mu 
\frac{{\hat k}_1-{\hat \kappa}_1+m}{-2k_1\kappa_1}
{\hat \epsilon}_1
u_1,
\end{eqnarray}
where $V^2=\kappa^2=(\kappa_1+\kappa_2)^2$.

\subsection{Matrix elements in leading approximation}\label{matrixel}

There are four kinematical regions contributed to 
the cross section in leading approximation: when one of the photon is observed and another in   
so-called $s$- and $p$-peaks. For $s$-peak ($p$-peak) the additional 
unobserved photon is emitted in the direction of the initial (final) lepton. Therefore, 
\begin{eqnarray}
\bigr(\sum_{i=1}^6{\cal M}_i\bigr)^2 ={\cal M}_{1s}^2+{\cal M}_{1p}^2+{\cal M}_{2s}^2+{\cal M}_{2p}^2,
\end{eqnarray}
where indices correspond to the unobserved photon, e.g., $1s$ means that the photon with momentum $\kappa_1$ is unobserved and in the $s$-peak.  

 The matrix element ${\cal M}_{1s}^2$ squared in the leading approximation is calculated assuming 
that the momentum $\kappa_1$ of unobserved photon is approximated as
\begin{equation}\label{sappro}
\kappa_1=(1-z_1)k_1.
\end{equation}
However this approximation have to be carefully applied after analyzing the the structure of poles, i.e., powers of $k_1\kappa_1$ in denominators. Only terms with the first-order pole ($1/k_1\kappa_1$) contribute the the cross section in the leading approximation. The second-order poles appear in the form of $m^2/(k_1\kappa_1)^2$ and does not contain the leading log after integration and taking the limit $m\rightarrow 0$. Only $J_{1\mu}$ and $J_{6\mu}$ have the pole, 
\begin{eqnarray}
J_{1\mu}&\approx &\frac{k_1\epsilon_1}{2\;\; k_1\kappa_2\;\; k_1\kappa_1}
{\bar u}_2 \gamma_\mu (z_1 {\hat k}_1-{\hat \kappa}_2){\hat \epsilon}_2
u_1 
\nonumber \\\label{J6app}
J_{6\mu}&\approx &\frac{z_1\;\;k_1\epsilon_1}{2\;\; k_1\kappa_2\;\; k_1\kappa_1}
{\bar u}_2 {\hat \epsilon}_2({\hat k}_2+{\hat \kappa}_2)\gamma_\mu 
u_1 ,
\end{eqnarray}
while $J_{2\mu}$, $J_{3\mu}$, $J_{4\mu}$, and $J_{5\mu}$ do not. It means that the interference term $(J_{1\mu}+J_{6\mu})(J_{2\nu}+J_{3\nu}+J_{4\nu}+J_{5\nu})^{\dagger}$ has the pole and therefore contributes to leading log approximation and that $J_{1\mu}+J_{6\mu}$ squared can have the pole of the second order. 

Calculating the interference, the equations (\ref{J6app}) can be applied, and the substitutions (\ref{sappro}) and $m=0$ can be used everywhere except in $k_1\kappa_1$ in the denominator. 
Both $J_{1\mu}$ and $J_{6\mu}$ are proportional to $k_1\epsilon_1$. Therefore, to calculate the interference one needs to calculate $\sum J_{i\mu} k_1\epsilon_1$  ($i=2,\dots ,5$) by averaging over unobserved photon polarization states. This results in
\begin{eqnarray}
\sum k_1\epsilon_1 (J_{3\mu}+J_{4\mu}) &= &
-\frac{{\bar u}_2 {\hat \epsilon}_2({\hat k}_2+{\hat \kappa}_2)\gamma_\mu 
u_1}{2(1-z_1)\;k_2\kappa_2},
\nonumber \\ 
\sum k_1\epsilon_1 (J_{2\mu}+J_{5\mu}) &= &
\frac{
{\bar u}_2 \gamma_\mu (z_1 {\hat k}_1-{\hat \kappa}_2){\hat \epsilon}_2
u_1}{2\; k_1\kappa_2\;z_1(1-z_1)}. 
\end{eqnarray}
and therefore
\begin{eqnarray}
&&(J_{1\mu}+J_{6\mu})(J_{2\nu}+J_{3\nu}+J_{4\nu}+J_{5\nu})^{\dagger}+h.c.=
\nonumber \\&&
=J^{BH}_\mu(z_1k_1,k_2)(J^{BH}_\nu(z_1k_1,k_2))^\dagger\frac{2}{k_1\kappa_1\;(1-z_1)}.
\end{eqnarray}

Calculating the term with $J_{1\mu}+J_{6\mu}$ squared, the poles have to be extracted in the form of $1/k_1\kappa_1$ and $m^2/(k_1\kappa_1)^2$, and only then the substitutions (\ref{sappro}) and $m=0$ can be used everywhere except in $k_1\kappa_1$ in the denominator. This results in
\begin{eqnarray}
&&(J_{1\mu }+J_{6\mu })
(J_{1\nu }+J_{6\nu })^{\dagger}|_{\kappa_1\to (1-z_1)k_1} 	
\nonumber \\&&\qquad
=J^{BH}_\mu(z_1k_1,k_2)(J^{BH}_\nu(z_1k_1,k_2))^\dagger\frac{1-z_1}{z_1\;k_1\kappa_1}.
\end{eqnarray}
We finally have in leading approximation:
\begin{eqnarray}\label{Jspeak}
&&M_{1s}^2
=\frac{4\pi\alpha }{k_1\kappa_1}{\cal M}^2_{BH}(z_1k_1,k_2)\frac{1+z_1^2}{z_1(1-z_1)}.
\end{eqnarray}

In the case when the unobserved photon is emitted parallel to the final electron the scalar product $k_2\kappa_1$ is small. For this case it is assumed that $\kappa_1=(z_2^{-1}-1)k_2$ resulting in    
\begin{eqnarray}\label{Jppeak}
&&M_{1p}^2
=\frac{4\pi\alpha }{k_2\kappa_1}{\cal M}^2_{BH}\left(k_1,\frac{k_2}{z_2}\right)\frac{1+z_2^2}{1-z_2}.
\end{eqnarray}

\subsection{Phase space and shifted kinematics}\label{phasespace}

\begin{figure}\centering
\scalebox{0.36}{\includegraphics{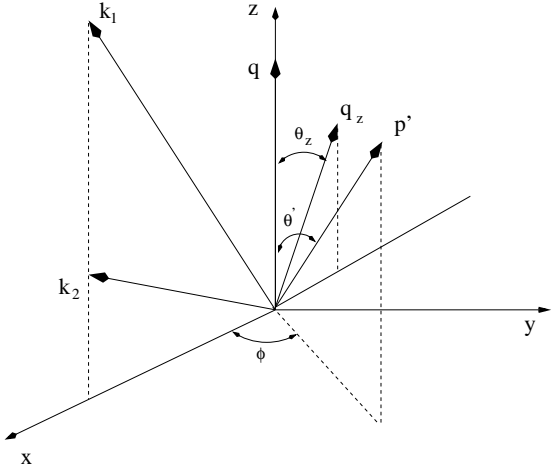}}
\caption{\label{figshifted}Definitions of vectors and angles in the Lab. frame}
 \end{figure}

The photon four-vectors appear in denominators of (\ref{Jspeak}) and (\ref{Jppeak}) in the form of scalar products $k_1\kappa_1$ and $k_2\kappa_1$. Two integrals over phase space of two photons are:
\begin{eqnarray}
&&\int \frac{d^3\kappa_1}{2\omega_1}\frac{d^3\kappa_2}{2\omega_2}
\frac{\delta(\Lambda-\kappa_1-\kappa_2)}{k_1\kappa_1}=
\nonumber\\&&
\qquad\qquad=\int \frac{d^3\kappa_1}{2\omega_1}
\frac{\delta(\Lambda^2-2\Lambda\kappa_1)}{k_1\kappa_1}
=\frac{\pi L}{w},
\nonumber\\\label{intL2}
&&\int \frac{d^3\kappa_1}{2\omega_1}\frac{d^3\kappa_2}{2\omega_2}
\frac{\delta(\Lambda-\kappa_1-\kappa_2)}{k_2\kappa_1}
\nonumber\\&&
\qquad\qquad=\int \frac{d^3\kappa_1}{2\omega_1}
\frac{\delta(\Lambda^2-2\Lambda\kappa_1)}{k_2\kappa_1}
=\frac{\pi L}{u},
\end{eqnarray}
where $\Lambda=k_1+p-k_2-p'$, $w=2k_1\Lambda$, $u=2k_2\Lambda$. Only terms containing the large (or leading) logarithm $L$ are kept. The results (\ref{intL2}) are immediately obtained if to consider the system of center-of-mass of two photons (${\boldsymbol\Lambda}=0$) with z-axis directed along ${\bf k}_1$ and ${\bf k}_2$, respectively.

The phase space of the final proton is parametrized as:
\begin{equation}
\frac{d^3p'}{2p'_0}=\frac{\sqrt{\lambda_t}}{8M^2}dtd\phi d\cos\theta '=
\frac{dt d\phi dV^2}{4\sqrt{\lambda _Y}},
\end{equation}
where  $V^2=\Lambda^2$ and $\theta '$ is the angle between $\bf q$ and $\bf p'$ . The relation 
\begin{equation}\label{VV2}
V^2=\frac{tS_x}{2M^2}+t-Q^2+
{\sqrt{\lambda_Y}\sqrt{\lambda_t}\over 2M^2}\cos\theta '
\end{equation}
was used to obtained the parametrization in terms of $V^2$. Finally the integration over $d\Gamma$ is
\begin{eqnarray}
\int \frac{d\Gamma}{k_1\kappa_1}&=&d\Gamma_0\frac{L}{8\pi^2w}dV^2,
\nonumber \\
\int \frac{d\Gamma}{k_2\kappa_1}&=&d\Gamma_0\frac{\pi L}{8\pi^2u}dV^2. 
\end{eqnarray}

The matrix elements squared for $s$- (and $p$-) peak contributions in eqs. (\ref{Jspeak}) and (\ref{Jppeak}) are expressed in terms of $z_1$ and $z_2$, therefore the variable $V^2$ (and $\cos \theta '$) has to be related to these variables. The equation for establishing this relation is obtained from condition in $\delta$-function argument of intermediate expressions in (\ref{intL2}) if to use the representation for $\kappa_1$ used in subsection \ref{matrixel}, i.e., $\kappa_1=(1-z_1)k_1$ for $s$-peak and $\kappa_1=(z_2^{-1}-1)k_2$ for $p$-peak. Below 
for representation of this equation and its solution we use the generalized notation included both $z_1$ and $z_2$. Substitution $z_2=1$ ($z_1=1$) has to be used to formally extract $s$-peak ($p$-peak) contribution. Also we define the 4-vector $q_z$: $q_z=z_1k_1-k_2$ for $s$-peak, $q_z=k_1-z_2^{-1}k_2$ for $p$-peak, or $q_z=z_1k_1-z_2^{-1}k_2$ in the generalized notation. Meaning of the used vectors is clarified in Figure \ref{figshifted}. Vector $q_z$ has meaning of ``true'' transferring momentum in the case of additional photon emitted. The vector is in the plane OXZ, its projection into OX and OZ axes are always negative and positive respectively. Its magnitude is always less than that of $q$.
The equation for establishing the relation between $z_1$, $z_2$, and $V^2$ in terms of introduced notation reads: 
\begin{eqnarray}
\Lambda^2-2\Lambda (q-q_z)
&=&\frac{\sqrt{\lambda_t\lambda_{Yz}}}{2M^2}
({\cos\bar\theta}-A)=0,
\label{Eqdelta}
\end{eqnarray}
where ${\cos\bar\theta}$ is the angle between ${\bf q}_z$ and $\bf p'$. It is expressed in terms of the angle between $\bf q$ and ${\bf q}_z$ (denoted by $\theta_z$) as
\begin{equation}\label{thetabar}
{\cos\bar\theta}=\cos\theta '\cos\theta_z-\sin\theta '\sin\theta_z\cos\phi ,
\end{equation}
where 
sinus and cosine of $\theta_z$ and the quantity $A$ are defined by kinematics in terms of $z_{1,2}$ and measured quantities:
\begin{eqnarray}
\cos\theta_z&=&{S_x(z_1S-z_2^{-1}X)+2(z_2^{-1}+z_1)M^2Q^2 \over \sqrt{\lambda_Y}\sqrt{\lambda_{Yz}}}, 
\nonumber \\
\sin\theta_z&=&{2(z_2^{-1}-z_1)MQ(SX-M^2Q^2)^{1/2} \over \sqrt{\lambda_Y}\sqrt{\lambda_{Yz}}} ,
\nonumber \\
A&=&-{(z_1S-z_2^{-1}X)t+2M^2(t-z_1z_2^{-1}Q^2)\over \sqrt{\lambda_t}\sqrt{\lambda_{Yz}}} ,
\nonumber \\\label{lyzz1z2}
\lambda_{Yz}&=&(z_1S-z_2^{-1}X)^2+4M^2z_1z_2^{-1}Q^2.
\end{eqnarray}
Eq. (\ref{Eqdelta}) has unique solution in the kinematically allowed region:
\begin{eqnarray}
\cos\theta '&=&\frac{A\cos\theta_z+\sqrt{{\cal D}_0}\sin\theta_z\cos\phi}{\cos^2\theta_z+\sin^2\theta_z\cos^2\phi},
\nonumber \\
\sin\theta '&=&\frac{\cos\theta_z\sqrt{{\cal D}_0}-A\sin\theta_z\cos\phi}{\cos^2\theta_z+\sin^2\theta_z\cos^2\phi},
\nonumber \\
\label{DD0}
{\cal D}_0&=&\cos^2\theta_z+\sin^2\theta_z\cos^2\phi-A^2.
\end{eqnarray}
 The direction of ${\bf q}_z$ defines new polar ($\bar\theta$) and azimuthal ($\bar\phi$) angles of the final proton and thus generates so-called shifted kinematics. The angle ($\bar\theta$) was defined in (\ref{thetabar}) and the angle $\bar\phi$ is related to measured $\phi$ as   
\begin{equation}\label{cosphiz}
\cos{\bar\phi}\sin{\bar \theta }=\cos\theta_z\sin\theta '\cos\phi+\sin\theta_z\cos\theta '
\end{equation}
and
\begin{equation}\label{sinphiz}
{\sin{\bar\phi}}\sin{\bar \theta }=\sin\theta '\sin\phi .
\end{equation}
The origin of the equation (\ref{sinphiz}) is clear because the projection of $\bf p'$ on axis OY is the same for original and shifted kinematics.
Recall, that the equations (\ref{Eqdelta}-\ref{sinphiz}) are used for both $s$- and $p$-peaks.  In the first case one sets $z_2=1$ and $z_1=1$ is set for the second case.  

The target polarization in shifted kinematics is calculated using the orthogonal transformation: 
\begin{eqnarray} 
{\bar\eta}_x&=&\cos\theta_z\eta_x+\sin\theta_z\eta_z, 
\nonumber \\
{\bar\eta}_y&=&\eta_y, 
\nonumber \\
{\bar\eta}_z&=&-\sin\theta_z\eta_x+\cos\theta_z\eta_z.
\end{eqnarray}
Thus, in the shifted kinematics the target polarization are not longer pure longitudinal or transversely polarized, therefore, the scalar products in shifted kinematics are then calculated using eqs. (\ref{etaxyz}).  

The variable $V^2$ is related to $z_{1,2}$ through eq. (\ref{VV2}) where $\cos\theta '$ is given by (\ref{DD0}) and quantities in the R.H.S. of (\ref{DD0}) depend on $z_{1,2}$ in (\ref{lyzz1z2}). Tedious, but straightforward calculation gives
\begin{eqnarray}
\frac{1}{w}\frac{dV^2}{dz_1}&=&-\frac{\sqrt{\lambda_Y}}{\sqrt{\lambda_{Yz}}} \frac{\sin\theta '}{\sqrt{{\cal D}_0}} , 
\nonumber \\ \label{dVV22}
\frac{1}{u}\frac{dV^2}{dz_2}&=&-\frac{\sqrt{\lambda_Y}}
{z_2^2\sqrt{\lambda_{Yz}}} 
\frac{\sin\theta '}{\sqrt{{\cal D}_0}}.
\end{eqnarray}
The R.H.S. of the equations (\ref{sinphiz},\ref{dVV22}) are taken using respective peak kinematics.

The equation for minimal value for $z_1$ (denoted by $z_{1}^m$) allowed by kinematics is $\cos\theta_z=A$. It follows from (\ref{VV2}):  
$V^2_{max}=tS_x/2M^2+t-Q^2+\sqrt{\lambda_Y}\sqrt{\lambda_t}/2M^2$ 
(that corresponds to $\cos\theta '=1$). The solution is 
\begin{equation}
z_{1}^m={Xt-2M^2t+\xi(XS_x-2M^2Q^2) \over St-2M^2Q^2+\xi(SS_x+2M^2Q^2)}.
\end{equation}
where $\xi^2=\lambda_t/\lambda_Y$. Similarly, for kinematics of $p$-peak we obtain 
\begin{equation}
z_{2}^m={Xt+2M^2Q^2+\xi(XS_x-2M^2Q^2) \over St+2M^2t+\xi(SS_x+2M^2Q^2)}.
\end{equation}

Both $z_{1}^m$ and $z_{2}^m$ do not depend on $\phi$.

\subsection{The lowest order RC to BH cross section}
\begin{figure}[t]
\centering
\scalebox{0.2}{\includegraphics{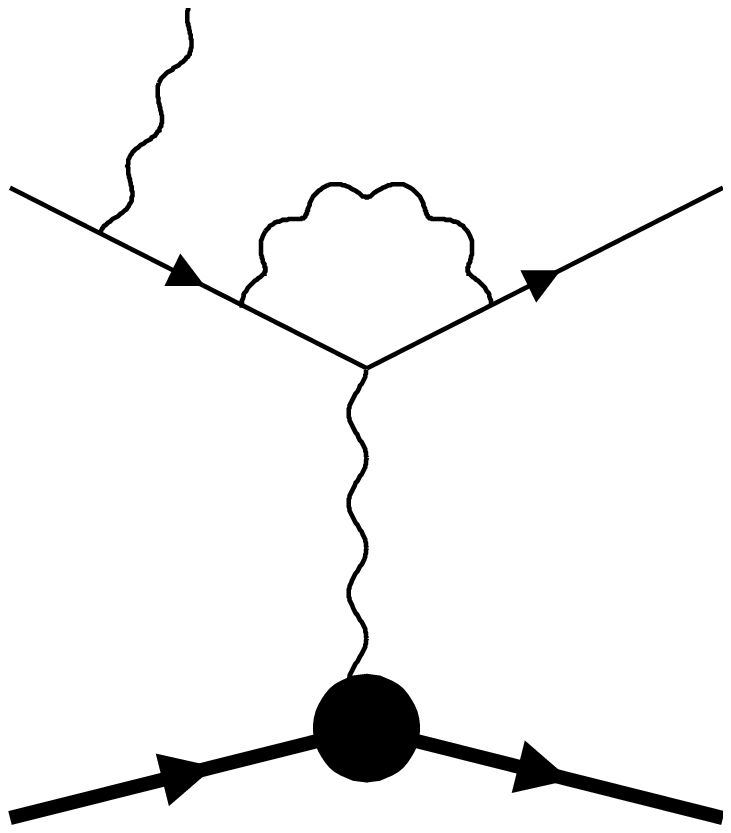}}
\hspace{0.25cm}
\scalebox{0.2}{\includegraphics{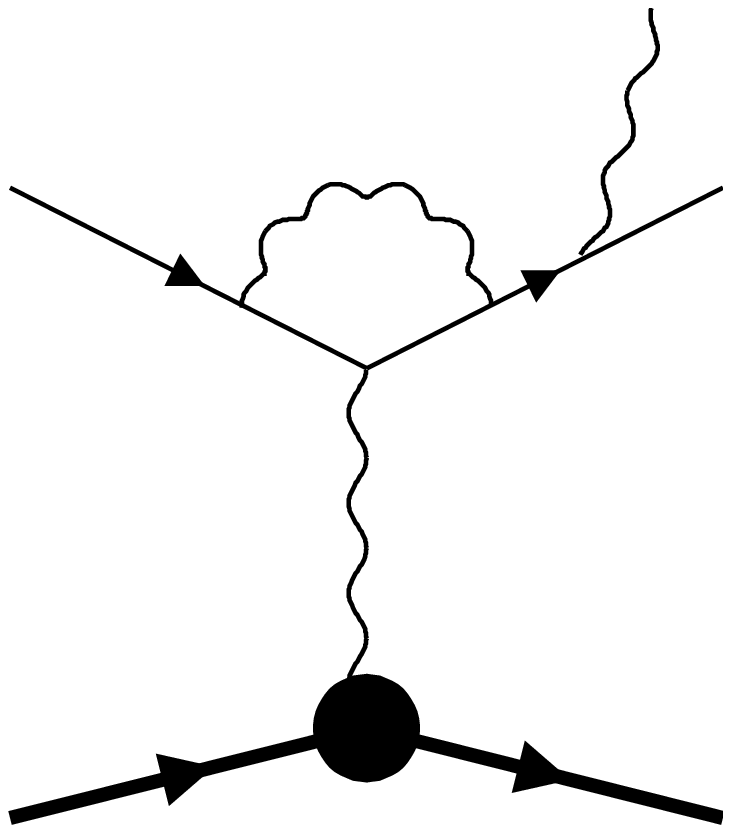}}
\hspace{0.25cm}
\scalebox{0.2}{\includegraphics{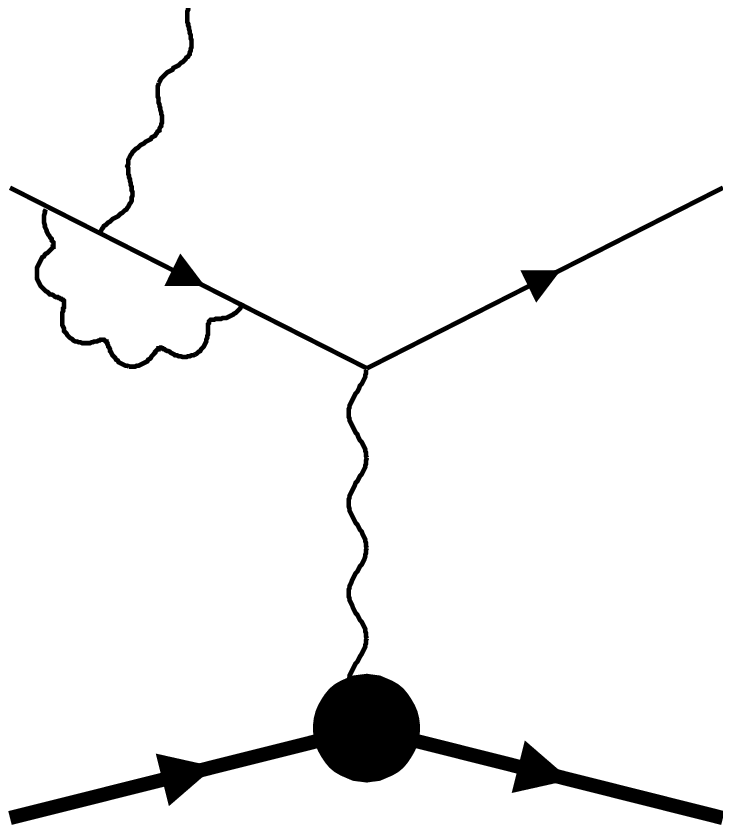}}
\hspace{0.25cm}
\scalebox{0.2}{\includegraphics{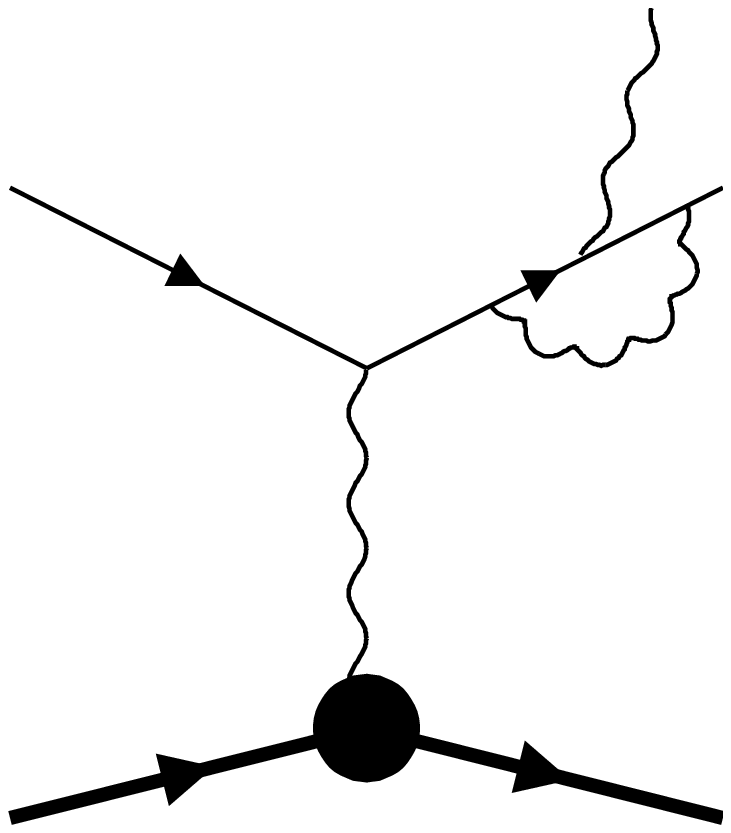}}
\\[-0.1cm]
{\bf \small a) \hspace{1.52cm} b) \hspace{1.52cm} c)\hspace{1.52cm} d)}
\\[0.1cm]
\scalebox{0.2}{\includegraphics{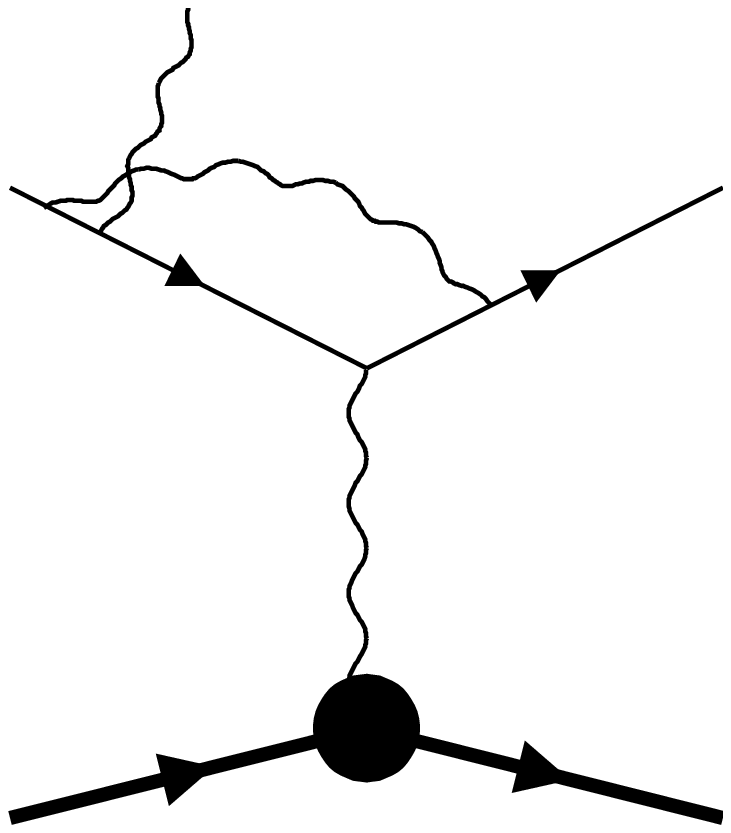}}
\hspace{0.25cm}
\scalebox{0.2}{\includegraphics{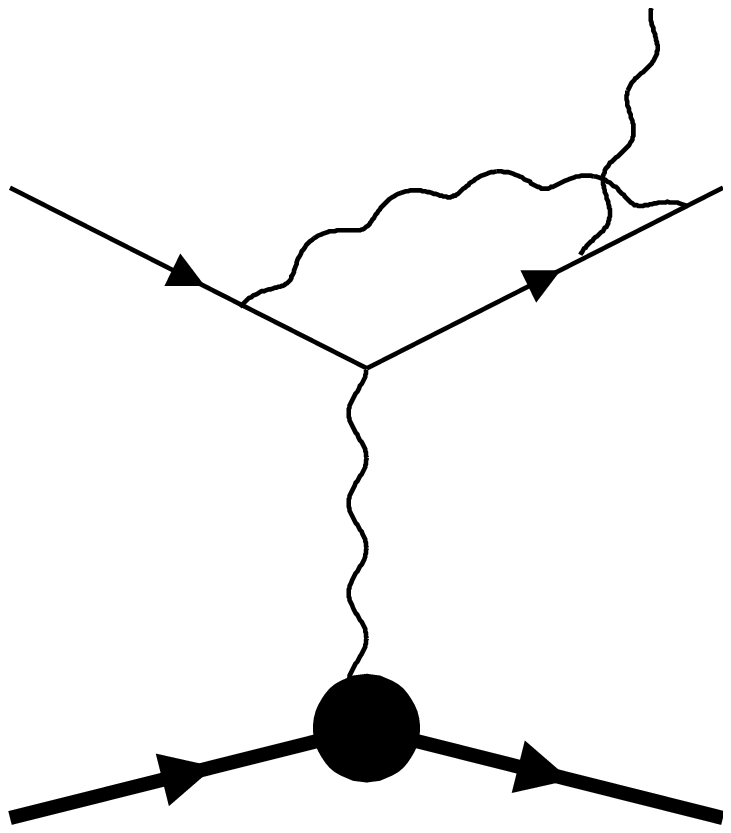}}
\hspace{0.25cm}
\scalebox{0.2}{\includegraphics{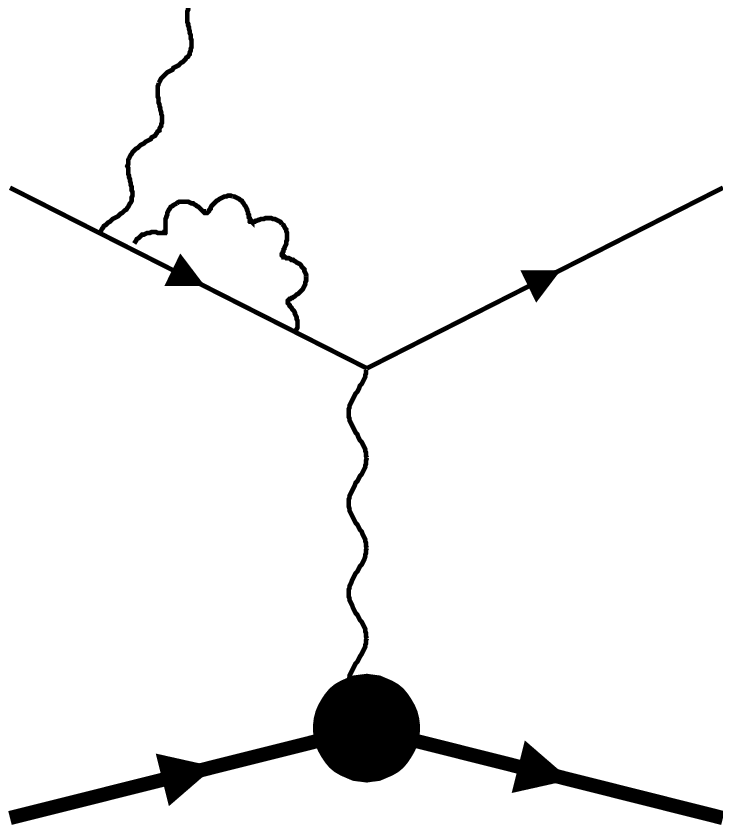}}
\hspace{0.25cm}
\scalebox{0.2}{\includegraphics{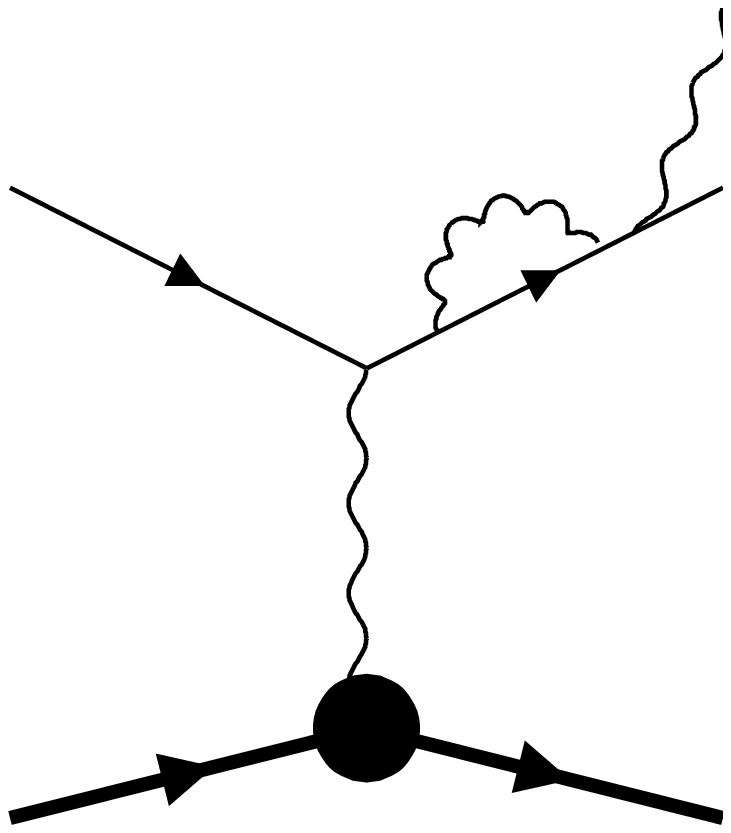}}
\\[-0.1cm]
{\bf \small e) \hspace{1.52cm} f) \hspace{1.52cm} g)\hspace{1.52cm} h)}
\\[0.1cm]
\scalebox{0.18}{\includegraphics{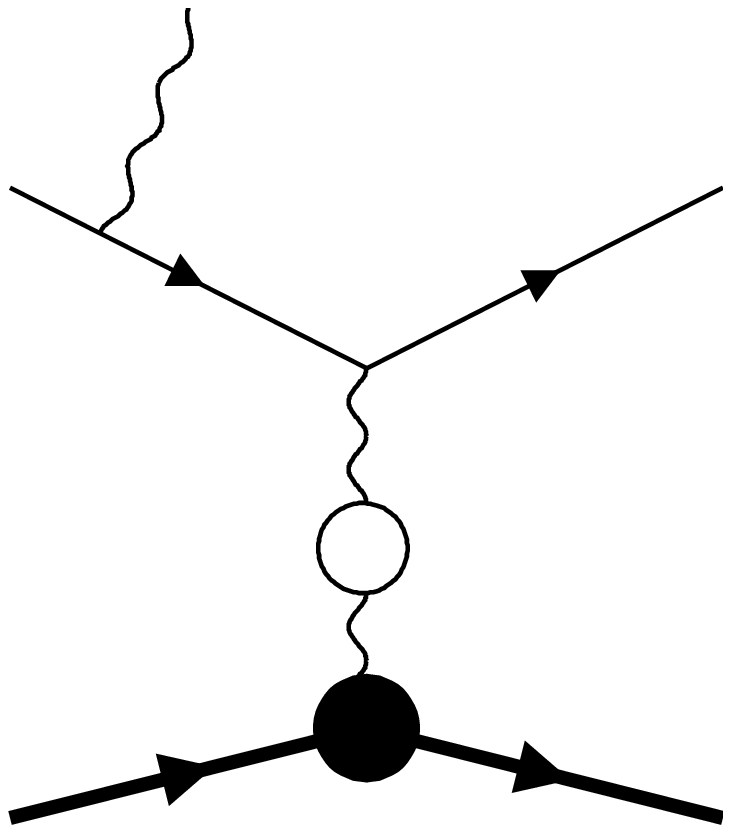}}
\hspace{0.25cm}
\scalebox{0.18}{\includegraphics{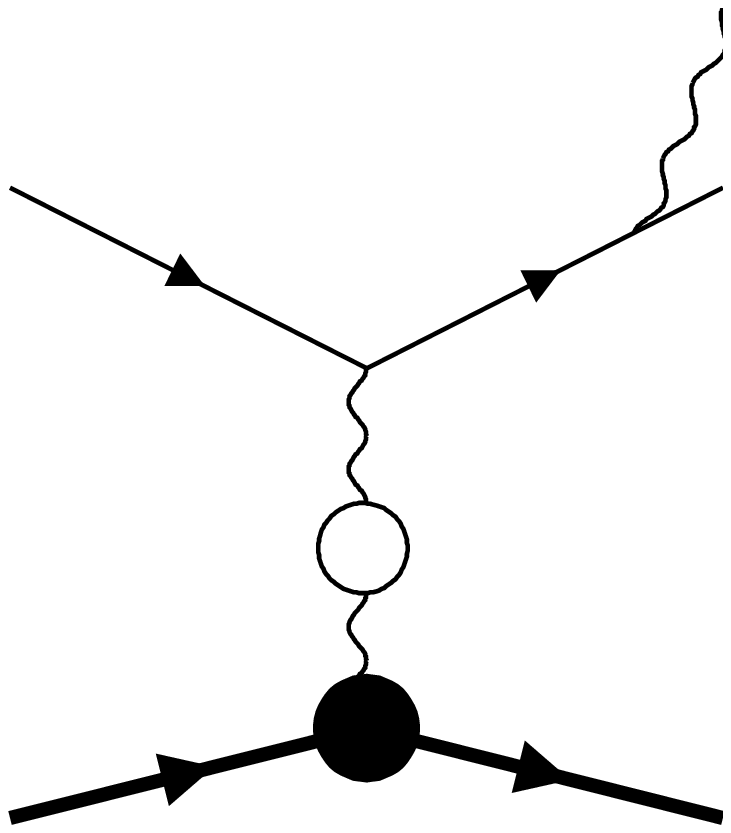}}
\\[-0.1cm]
{\bf \small i) \hspace{1.5cm} j)}
\caption{\label{VVgraphs}Feynman graphs of one-loop effects for the BH cross section}
 \end{figure}

 Combining results obtained in Sections \ref{matrixel} and \ref{phasespace} we find the cross section of two photon emission as:
\begin{eqnarray}
&&\sigma _{s}(S,x,Q^2,t,\phi)=\frac{\alpha}{2\pi }L \times
\nonumber\\ && \qquad
\int\limits_{z_{1}^m}^1 dz_1\frac {1+z_1^2}{1-z_1} 
{\sin\theta_s ' \over {{\cal D}_{0s}^{1/2}}}
 \left(\frac{ x_s}{x}\right)^2\sigma _{BH}(z_1S,x_s,z_1Q^2,t,\bar\phi_s),
\nonumber\\&&\label{rvkladp}
\sigma _{p}(s,x,Q^2,t,\phi)=\frac{\alpha}{2\pi }L \times
\\&& 
\int\limits_{z_{2}^m}^1 dz_2\frac {1+z_2^2}{z_2(1-z_2)}
{\sin\theta_p ' \over {{\cal D}_{0p}^{1/2}}}
\left( \frac{ x_p}{x}\right )^2\sigma _{BH}(S,x_p,z_2^{-1}Q^2,t,{\bar \phi}_p),\nonumber
\end{eqnarray}
where $x_s=z_1Q^2/(z_1S-X)$ and $x_p=Q^2/(z_2S-X)$ are Bjorken $x$ in shifted kinematics; $\sin\theta '$ and $\bar\phi$ are given by (\ref{DD0}) and (\ref{sinphiz})---subscript explicitly indicates the type of kinematics for that these quantities have to be calculated.

The integrals in (\ref{rvkladp}) are divergent at upper integration limit, therefore it is regularized using a parameter $\omega_{min}$ separating the integration region on the part corresponding to emission of soft and hard photons. For $z_{1,2}$ the regulating parameter $\Delta$ is $\Delta=\Delta_1=2M\omega_{min}/S$ for $s$-peak and $\Delta=\Delta_2=2M\omega_{min}/X$ for $p$-peak.

The contributions of loops (Fig. \ref{VVgraphs}{\it a-h}) and soft photon emission are known \cite{ByKuTo2008PRC}. Their sum is proportional to BH cross section
\begin{equation}
\sigma_V=\frac{\alpha}{\pi}\biggl(\log\frac{4M^2\omega_{min}^2}{SX}+\frac{3}{2}\biggr)
L\sigma_{BH}
\end{equation}
and can be presented as
\begin{equation}
\label{vvklad}
\sigma_V=-\frac{\alpha L}{2\pi}\sigma_{BH}\biggl(
\int\limits_{0}^{1-\Delta_1} dz_1 \frac{1+z_1^2}{1-z_1} 
+\int\limits_{0}^{1-\Delta_2} dz_2 \frac{1+z_2^2}{1-z_2}\biggr). 
\end{equation}
Sum of (\ref{rvkladp}) and (\ref{vvklad}) is infrared free and regularization can be removed: $\Delta_{1,2}=0$. The result for the observed cross section is: 
\begin{widetext}
\begin{eqnarray} 
&&\sigma _{obs}^{1-loop}(S,x,Q^2,t,\phi)=(1 + 2\Pi (t))\sigma _{BH}(S,x,Q^2,t,\phi)
+ \frac{\alpha}{2\pi }L
\Biggl [\;
\nonumber
\\[0.1cm]&&
\quad\int\limits_{0}^1 dz_1\left (\frac {1+z_1^2}{1-z_1} \right )
\left(
{\sin\theta_s '   \over {{\cal D}_{0s}^{1/2}}  }
\theta(z-z_1^m) \left(\frac{x_s}{x}\right )^2\sigma _{BH}(z_1S,x_s,z_1Q^2,t,{\bar\phi}_s)
-
\sigma _{BH}(S,x,Q^2,t,\phi)\right)
\nonumber\\[0.1cm]&& +
\int\limits_{0}^1 dz_2\left (\frac {1+z_2^2}{1-z_2} \right )
\left( 
{\sin\theta_p '  \over {{\cal D}_{0p}^{1/2}}}
\theta(z-z_2^m)\frac 1{z_2}\left (\frac{x_p}{x}\right )^2
\sigma _{BH}(S,x_p,z_2^{-1}Q^2,t,{\bar\phi}_p)-
\sigma _{BH}(S,x,Q^2,t,\phi)\right)
\Biggr ].
\label{1l}
\end{eqnarray}
\end{widetext}
Here $ \Pi (t)=\alpha/(2\pi)\delta_{vac}$  and $\delta_{vac}$ is the contribution of vacuum polarization by leptons and hadrons (Fig. \ref{VVgraphs}{\it i,j}) calculated as in \cite{AKSh1994} (see eq. (21) and discussion before eq. (20)).

Behavior of the cross section for $t$ close to kinematical bounds (i.e., in the region where $t\sim t_1$ and $t\sim t_2$) deserves special attention. The integrals in (\ref{1l}) become infinite when $t \rightarrow t_1$ or $t \rightarrow t_2$. In this limit $z_1^m=1$ and $z_2^m=1$. To extract the divergence  the part of integrals in (\ref{1l}) from 0 to $z_1^m$ or $z_2^m$ need to be calculated analytically resulting in:
\begin{equation}
\sigma _{obs}^{1-loop}=\Bigl( 1 + \frac{\alpha}{\pi}\bigl( \delta_{vac}+\delta_{inf}+\delta_{fin}\bigr)\Bigr) \sigma _{BH} + \sigma_{F}.
\end{equation}
where $ \sigma_{F}$ is the non-divergent contributions of remaining integrals (i.e., as in (\ref{1l}), but  with low limits $z_1^m$ and $z_2^m$). The correction terms
\begin{eqnarray}
\delta_{fin}&=&\frac{L}{4}\bigl(z_1^m(2+z_1^m)+z_2^m(2+z_2^m)\bigr), \\
\delta_{inf}&=&L\bigl( \log(1-z_1)+\log(1-z_2)\bigr). \nonumber
\end{eqnarray}
represent the the finite and infinite parts of the results of the analytical integration. The source of occurrence of the divergence is known \cite{YennieFrautschiSuura1961}. The divergence is canceled by taking into account multiple soft photon emission. We follow the so-called exponentiation procedure suggested in \cite{Shumeiko}:
\begin{eqnarray}\label{exponentiation}
&&\Bigl( 1 + \frac{\alpha}{\pi}\bigl( \delta_{vac}+\delta_{inf}+\delta_{fin}\bigr)\Bigr)
\rightarrow \\
&&\qquad \qquad \exp\bigl(\frac{\alpha}{\pi} \delta_{inf}\bigr)\Bigl( 1 + \frac{\alpha}{\pi}\bigl( \delta_{vac}+\delta_{fin}\bigr)\Bigr). \nonumber
\end{eqnarray}
After this procedure the observed cross section vanishes at the kinematical bounds on $t$.

\subsection{Higher order corrections}

In previous section we found RC to BH cross section in leading approximation
induced by lepton leg  in the lowest order over $\alpha $. 
The generalization of eq. (\ref{1l}) on highest order over $\alpha $
using electron structure function method
suggested in \cite{KurFad85} (see also \cite{ESFRAD,ESF})
has a form:
\begin{eqnarray} 
&&\sigma _{obs}(S,x,Q^2,t,\phi)=
\int\limits_{z_1^m}^1 dz_1
\int\limits_{z_{2,1}^m}^1 \frac{dz_2}{z_2}
D(z_1,Q^2)D(z_2,Q^2)
\nonumber \\&&\qquad\times
\left (\frac {x_{sp}}x\right )^2
{\sin \theta '  \over {{\cal D}_{0}^{1/2}}}
\hat{\sigma } _{BH}(z_1S,x_{sp},z_2^{-1}z_1Q^2,t,{\bar\phi}),
\label{hl}
\end{eqnarray}
where ${\hat \sigma} _{BH}=
\sigma _{BH}[\alpha^3 \to \alpha^3/(1-\Pi(t))^2]$,
$x_{sp}=z_1 Q^2/(z_1 z_2S-X)$
and 
\begin{equation}
z_{2,1}^m={Xt+2z_1M^2Q^2+\xi(XS_x-2M^2Q^2) \over z_1 St+2M^2t+z_1\xi(SS_x+2M^2Q^2)}.
\end{equation}

The electron structure function $D(z,L)$ includes contributions due to
photon emission and pair production
\begin{equation}\label{4, unpol.  EST}
D = D^{\gamma} + D^{e^+e^-}_N + D^{e^+e^-}_S \ ,
\end{equation}
where $D^{^{\gamma}}$ is responsible for the photons radiation and
$D^{^{e^+e^-}}_N $ and $D^{^{e^+e^-}}_S$ describe pair production in
non-singlet (by single photon mechanism) and singlet (by double
photon mechanism) channels, respectively. The explicit expression for
$D(z,L)$ are given by eqs. (5-7) of ref. \cite{ESFRAD}.

Notice, that equation (\ref{1l}) can be reproduced by expansion of  (\ref{hl}) over $\alpha $ and keeping only zero and first order. 

\section{Numerical estimates}\label{SectNumeric}

Numerical analysis is designed to evaluate the RC for the cross section and the Fourier coefficients in the kinematics of modern measurements at Jlab \cite{Camacho_etal_2006_PRL,Mazouz_etal_2007_PRL,Girod_etal_2008_PRL}. Specific focus in this analysis will be on i) the $t$- and $\phi$-dependencies of the magnitude of RC factor and ii) investigation of RC for the Fourier coefficients both non-vanishing and vanishing at the level of the BH cross section.  

\subsection{Cross section}

\begin{figure}\centering
\scalebox{0.65}{\includegraphics{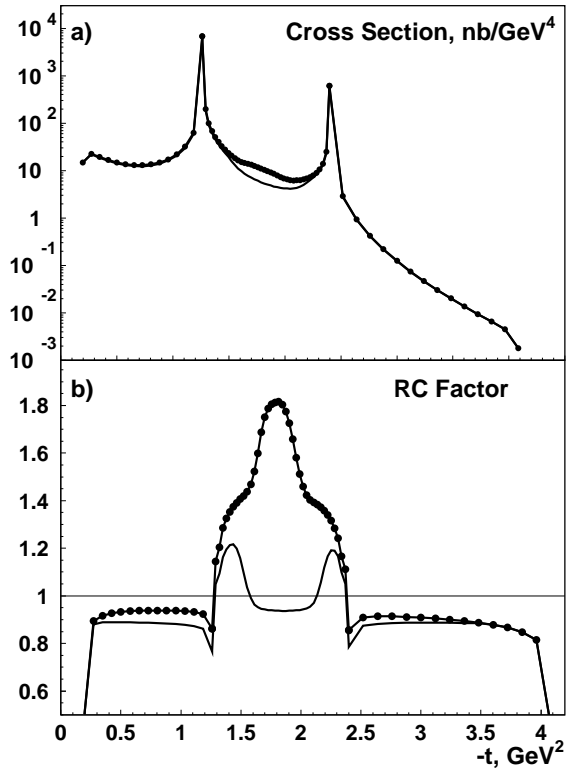}}
\caption{\label{RCF}The observed cross sections of the BH process (upper plot) and respective RC factors (lower plot) for beam energy 5.77~GeV, $x$=0.4, and $Q^2$=1.8GeV$^2$. The line with (without) dots shows the results of calculation without (with) the cut on missing energy ($E_\gamma<$0.3 GeV). }
\end{figure}

The $t$-distribution of the BH cross section has two sharp peaks that correspond to collinear radiation. The typical shapes of the $t$-dependence of the BH cross section with RC are represented in Figure \ref{RCF}a. Figure \ref{RCF}b gives $t$-dependence of the RC factor for the given kinematical points. The plots for spin dependent parts looks similar for both longitudinal and transverse polarizations (not shown). In this analyses the cross section integrated over $\phi$ is considered. 

Analyses of $t$-dependence presented in Figure \ref{RCF} revealed three specific regions in which the shapes of RC deserve attention and further clarification: i) the region close to bound over $t$ where RC factor goes rapidly down, ii) the region close to collinear peaks, and iii) the region between the peaks where RC factor can reach large values, however, capable of being suppressed by a cut on missing energy (also shown in Figure \ref{RCF} by the line without dots). 

Decrease of the RC factor in the region close to the bounds (i.e., $t\sim t_1$ or $t\sim t_2$) is  simply the reflection of the fact that observed cross section after the exponentiation procedure (\ref{exponentiation}), as well as the observed cross section (\ref{hl}) included higher order corrections, goes to zero at these kinematical bounds.

\begin{figure}\centering
\vspace{2mm}
\scalebox{0.65}{\includegraphics{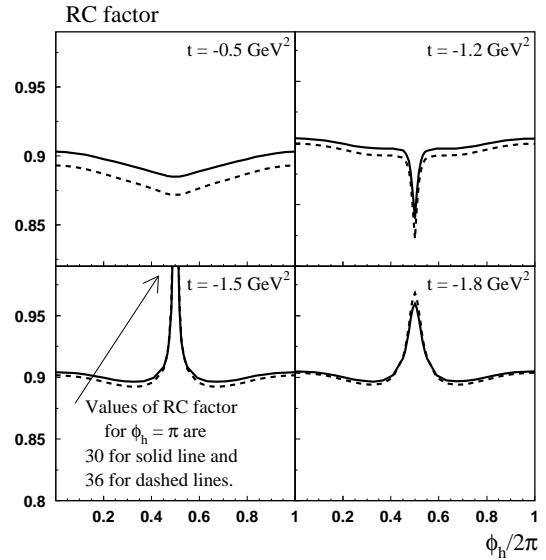}}
\caption{\label{RCFphidep}The RC factor vs. $\phi$  for beam energy 5.77~GeV, $x$=0.4, and $Q^2$=1.8GeV$^2$. The solid (dash) lines correspond the result of calculation  without (with) the cut on missing energy ($E_\gamma<$0.3 GeV).}
\end{figure}

In the region close to $s$- and $p$-peaks, i.e., when
\begin{equation}
t=t_s=-{Q^2 X \over S-Q^2}, \qquad 
t=t_p=-{Q^2 S \over X+Q^2}.
\end{equation}
the RC factor slowly decrease (when $-t$ approaches to the $s$-peak from the left or $p$-peak from the right, see Figure    
 \ref{RCF}),   reach its minimum at $t=t_s$ or $t=t_p$, and then rapidly increase, reach its maximum at $t=-Q^2$. Analysis of the integrand shows that the region around the point  
 $\phi=\pi$ is responsible for this difference. Therefore the $\phi$-dependence of the RC factor (Figure \ref{RCFphidep}) was analyzed. The RC factor typically has flat behavior except the point $\phi=\pi$ corresponding to the situation when the scattering and production planes coincide. In this case the RC factor can rapidly increase. Further analysis of the integrand showed that this increase of the RC factor is due to contribution of the second integral in (\ref{1l}) when $w_0$ is very small. The the second integral in (\ref{1l}) describes the $p$-peak contribution of the one photon, and the region of small $w_0$ corresponds to the $s$-peak of the second photon. Therefore, the large contribution comes from the two photon emission process when two irradiated photons are collinear to initial and final electrons. Corresponding BH process (i.e., one photon emission process) is the process with the emitted photon with 4-momentum corresponding to the sum of momenta of the two collinear photons. This photon is not collinear and therefore respective cross section of BH process is not large. The RC factor defined as the ratio of observed cross section (with large contribution of the two collinear photons) to the BH cross section (with not large BH cross section) can become larger than 2, i.e., RC to BH cross section can be larger than the BH cross section. Roughly the effect for RC factor can be estimated as $1+\alpha L^2$ (one collinear photon produce one leading log $L$). If $L\sim 15$ then the RC factor equals   2.64.

\subsection{RC and azimuthal structure of the cross section}

\begin{figure}\centering
\scalebox{0.75}{\includegraphics{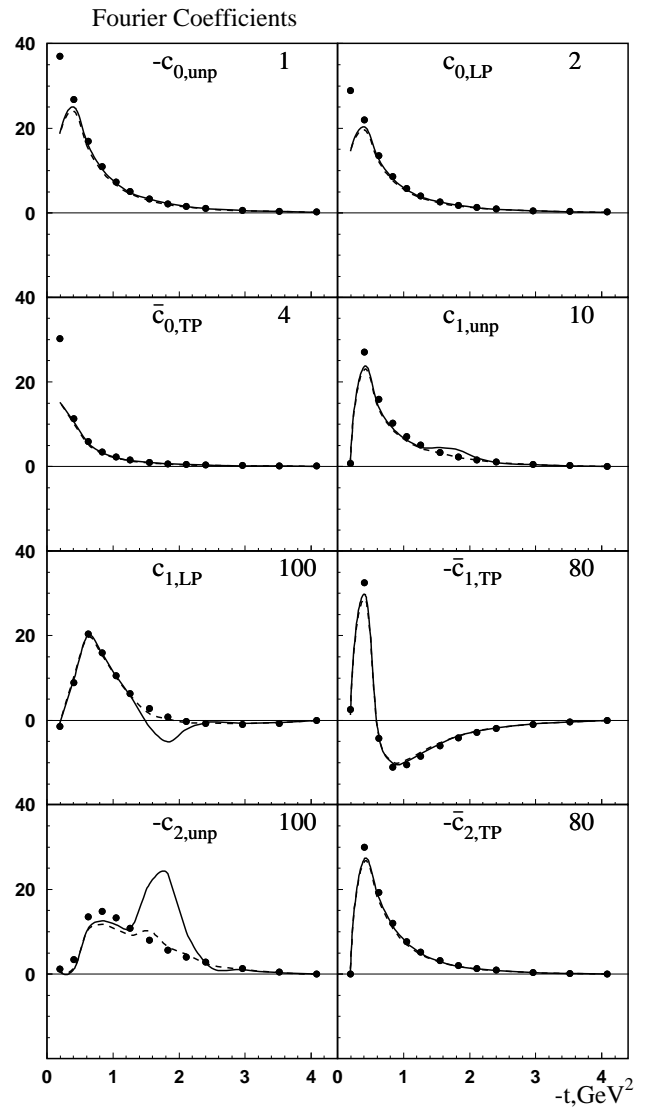}}
\caption{\label{RCFfour}The Fourier coefficient with nonzero contribution to the BH cross section (\ref{scc0}). Points represent the BH cross section, and solid (dashed) line represents the observed cross section (\ref{hl}) calculated without (with) the cut $E_\gamma=0.3 GeV$. The coefficients were rescaled to use the same scale on all plots and to compare them among each other: the original values of coefficients are calculated by dividing the values obtained from plot by the rescaled factor.}
\end{figure}

Azimuthal structure for the unpolarized BH cross section and for the longitudinally and transversely polarized cross sections are represented by Fourier coefficients defined in Section \ref{polar}. There are eight non-zero Fourier coefficients: three for unpolarized cross section, two for longitudinally polarized, and three for transversely polarized. Radiatively corrected Fourier coefficients are calculated using eqs. (\ref{cc0},\ref{barc}) with the observed cross section (\ref{hl}) substituted instead of $\sigma_{BH}$. Figure \ref{RCFfour} presents the results for these coefficients calculated using the BH and observed cross sections. The observed cross section was calculated with and without kinematical cut on maximal photon energy $E_\gamma=0.3 GeV$. One can see from this plot that the eight coefficients are quite stable in respect to RC. Similarly to the case of the cross section the regions with noticeable effect from RC are the region of small $-t$ and the region of $t$ close to (and between of) the $s$- and $p$-peaks. Also the results show that using the cut on missing energy suppresses the correction in the latter region.    

\begin{figure}\centering
\scalebox{0.7}{\includegraphics{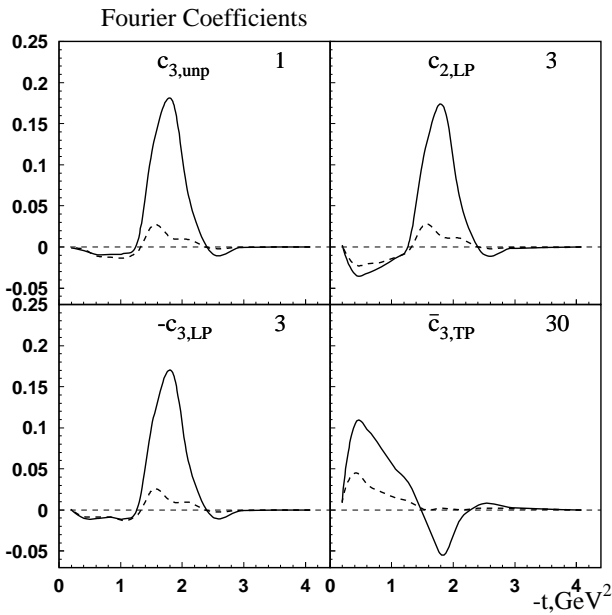}}
\caption{\label{RCFfour2}The Fourier coefficients not contributing to the BH cross section (\ref{scc0}).  Solid (dashed) line represents the observed cross section (\ref{hl}) calculated without (with) the cut $E_\gamma=0.3 GeV$. The definition of the rescaled factors are the same as in Figure \protect\ref{RCFfour}.}
\end{figure}

In contrast to the BH cross section, the azimuthal structure of the observed cross cannot be represented neither in terms of this eight coefficients nor in terms of any finite number of such coefficients. This is because of complicated and nonlinear dependence of the observed cross section on $\phi$. Several  coefficients representing next terms in the Fourier series are presented in Figure \ref{RCFfour2}. All of them are defined through $\cos (n\phi )$. The Fourier coefficients with $\sin (n\phi )$ were also investigated (all of them vanish at the level of the BH process), however no significant contributions at the level of observed cross section were found. 

\section{Discussion and Conclusion}\label{SectDiscussion}

In this paper we calculated RC to the BH cross section in leading approximation. Both unpolarized and polarized parts of the cross sections were considered. All final formulae are presented in analytical form. Details of calculation of matrix element squared are given with specific attention to occurrence of mass terms non-vanishing in the approximation of small lepton mass. Phase space was parametrized using the notion of shifted kinematics resulting in compact and convenient parametrization for the two-photon phase space and opportunities for analytic integration over angles. Numerical analysis of the effects of RC was focused on the RC to cross section and the Fourier coefficients representing the angular dependence of the BH and observed cross sections. 

Analysis of the RC to the BH cross section revealed the kinematical regions where RC can exceed the BH cross section in several times. This is the region with scattering azimuthal angle equaling $\pi$. The situation when both photons are collinear (one is collinear to initial lepton and another is collinear to final lepton) are kinematically allowable. Since the photon in respective BH process is not collinear (its momentum is the sum of two collinear photons), the BH cross section is not so large. As a result, the RC factor can be at the level of several dozens.  

There are eight Fourier coefficients contributing to the BH cross section with arbitrary polarized target, while new coefficients appear in the Fourier expansion at the level of the observed cross sections. The calculation of the contributions of the additional terms of the Fourier series can be significant. For example, as follows from the comparison of results in Figures \ref{RCFfour} and \ref{RCFfour2} the effect of "new" coefficient $c_{3,unp}$ can reach 10\% from the effect of main contribution represented by $c_{0,unp}$. This effect, however, can be suppressed by using the kinematical cut on missing energy. Note that experimental procedure of extraction of the Fourier coefficients from data is based on the fitting of observed cross section by the functions representing the angular structure of the BH cross section. Occurence of large terms of the next orders in Fuorier expansion can results in systematical uncertainties in kinematic regions where the effects of these additional terms is noticable.        

One feature of the calculation is that the lepton mass cannot be completely eliminated in the expressions for the BH cross section (\ref{dGamma}). First, the lepton mass has to be kept in the lepton propagators $w_0$ and $u_0$. Since the propagators are proportional to $E_L-p_Lcos\beta$ ($E_L$ and $p_L$ are energy and momentum of the lepton, and $\beta$ is the angle between momenta of the lepton and photon), there are kinematical points where $w_0$ or $u_0$ vanish in massless approximation making the BH cross section infinite. These points can be excluded when the BH process is investigated experimentally. However RC calculation requires integration of the BH cross section over  broad kinematical area and the singular point occur in the integration region. Therefore the lepton mass has to be kept in the expressions for $w_0$ and $u_0$. This is the reason of occurrence of the $m^2$ in (\ref{lambdauw}). Second, terms in the BH cross section containing $m^2$ in numerator and $w_0$ or $u_0$ squared in denominator are also infinite in massless approximation for certain $\phi$ and result in finite (independent of the lepton mass) terms after integration over $\phi$. Such terms were kept in the expression for the BH cross section (\ref{dGamma}). Note that our experience of dealing with RC tells us that such terms can give important contribution to the observed cross section (e.g., in DIS cross section measurements).    

The motivation for our calculation was the lack of complete calculations of the RC performed with accuracy to be controlled.  
One-loop correction and soft photon emission was calculated by Vanderhaeghen et al. \cite{Vanderhaeghen2000}.  Detailed consideration of one-loop correction was done. Box-type diagrams were evaluated in the style of ref. \cite{MaximonTjon2000PRC}. However, the radiative tail corresponding to photon emission processes was calculated in the approximation where the photon energy is very small compared to the lepton momenta.

Bytev, Kuraev, and Tomasi-Gustafsson \cite{ByKuTo2008PRC} applied the method of the electron structure functions to calculate RC due to two photon emission in  
 the process $e^-\mu^+$ that was chosen as a model process of DVCS. Main focus in this calculation was on the correction to the helicity-odd part of cross section, i.e., the interference between BH and DVCS amplitudes. Authors were focused on another experimental design: they integrated over the energy fraction of scattered electron.    

 The task of the calculation of RC to BH is closely related to the task of RC to radiative tail from the elastic peak that is the important (and often dominant) contribution to RC in DIS measurements. The radiative tail is simply the BH cross section integrated over photonic variables, i.e., over $\phi$ and $t$. Integration over $\phi$ is performed analytically, and because of dependence of the cross section on formfactors the integration over $t$ is left for numerical analysis. Programs for RC calculation of the radiative tail such as POLRAD 2.0  
  \cite{POLRAD20} include both the contribution of the radiative tail (correspondent to the BH cross section) and approximate calculation of the RC to the radiative tail (correspondent to two-photon emission and loop effects) \cite{AISh1998,AkKuSh2000PRDtail}. The approach to exact calculation of the RC to the radiative tail was developed by 
 Akhundov, Bardin, and Shumeiko \cite{AkBaSh1986YP}. They used the formalism of covariant extraction and cancellation of infrared divergence and calculated the QED corrections to the elastic radiative tail for unpolarized case. No analytical expressions represented the result of the exact calculation were published.

 The formulae in this paper are presented in analytical form providing good starting point for more precise calculations. One further generalization can be done using the approach of \cite{AkBaSh1986YP} to exactly calculate the lowest order correction to polarized BH cross section. Another direction for generalization is to apply the developed formalism for RC to DVCS, i.e., interference of  BH and DVCS amplitudes. Hadronic part of DVCS is known from refs. \cite{BKM2002,BeMu2010PRD}. 
 
 Note also that the developed formulae are obtained for the specific way of reconstruction of kinematic variables. Specifically, leptonic and hadronic momenta are used to reconstruct the kinematics of the BH process. Kinematical variables of the photon were assumed to be unmeasured. If information about photonic variables are involved into reconstruction of the kinematics of the BH  
 process the calculation presented in this paper requires modification. Universal way to avoid multiple calculation to cover all possibilities for data analysis designs is the development of the Monte Carlo generator of the BH process with the additional process with two photons. Any specific choice of base set of kinematical variables can be used for this construction including those considered in this paper. 
 
 \medskip
 \noindent{\bf Acknowledgments}. The authors are grateful to Harut Avakian for
interesting discussions and comments. This work was supported by DOE contract No. DE- AC05-06OR23177, under which Jefferson Science Associates, LLC operates Jefferson Lab.
 
\bibliography{dvcsll}{}

\end{document}